\documentclass[preprint, prl]{revtex4}

\usepackage[latin1]{inputenc}                   
\usepackage{graphicx,color,textcomp}
\usepackage{natbib}

\begin{document}

\title{An expanding universe without dark matter and dark energy}

\author{Pierre Magain}

\affiliation{Institut d'Astrophysique et de Géophysique, Université de Liège, Bat. B5C, Allée du 6 Août, 17, B-4000 Liège 1, Belgium}

\email{Pierre.Magain@ulg.ac.be}

\begin{abstract}
Assuming that observers located inside the Universe measure a time flow which is different from the time appearing in the Friedmann-Lemaître equation, and determining this time flow such that the Universe always appears flat to these observers, we derive a simple cosmological model which allows to explain the velocity dispersions of galaxies in galaxy clusters without introducing dark matter.  It also solves the horizon problem without recourse to inflation.  Moreover, it explains the present acceleration of the expansion without any resort to dark energy and provides a good fit to the observations of distant supernov\ae.  Depending on the present value of the matter-energy density, we calculate an age of the Universe between 15.4 and 16.5 billion years, significantly larger than the 13.7 billion years of the standard $\Lambda$CDM model.  Our model has a slower expansion rate in the early epochs, thus leaving more time for the formation of structures such as stars and galaxies.
\end{abstract}


\maketitle

\section{Introduction}

General relativity (GR) is an extremely powerful theory of gravitation.  Since its introduction nearly a century ago \cite{Einstein1916}, it has been confirmed by a number of experiments and observations, most notably the perihelion precession of Mercury, the gravitational redshift and the deflection of light by massive bodies.

It has also been applied to describe the evolution of the Universe as a whole, bringing cosmology into the realm of scientific research.  In particular, the model which currently gives the best description of the large-scale structure and evolution of the Universe, namely the $\Lambda$ Cold Dark Matter ($\Lambda$CDM) model, is based on the equations of GR, including the cosmological constant $\Lambda$, first introduced then discarded by Einstein himself.

However, in order to provide an accurate description of large-scale and long-term phenomena, a number of artifacts had to be added to the theory.  The motion of stars in the outskirts of spiral galaxies and the velocity dispersions of galaxies in clusters required the addition of dark matter.  The acceleration of the expansion of the Universe called for the introduction of dark energy.  The so-called flatness and horizon problems were solved by postulating an early phase of dramatic expansion named inflation.

One may remark that all the successful tests of GR are quasi-instantaneous in terms of cosmological time.  Conversely, the dark matter, dark energy and inflation artifacts are added to bring the predictions of the models in agreement with observations dealing with very long term phenomena.  Indeed, it takes several hundred millions of years for stars to revolve around the center of their galaxy; it takes billions of years for galaxies to complete an orbit in clusters; the acceleration of expansion is also measured over billions of years; the flatness and horizon problems relate present-day observations to phenomena that occurred billions of years ago.

Thus, while GR is extremely successful in describing phenomena occuring on short time scales, it requires the addition of artifacts for explaining very long term phenomena.  It might therefore appear useful to investigate extensions of the theory which, while conserving its instantaneous properties, would modify its cosmological applications.

\section{Relativity and cosmic time}

Assuming that the Universe is homogeneous and isotropic on a large scale, Einstein's equations of general relativity can be reduced to the Friedmann-Lemaître (FL) equation for the evolution of the scale factor $R$ of the Universe:
\begin{equation}
\left[\left(\frac{1}{R}\frac{dR}{dt}\right)^2 - \frac{8 \pi G \rho}{3}\right] R^2 = -k c^2
\end{equation} 
where $G$ is the gravitational constant, $c$ the speed of light, $\rho$ the matter-energy density and $k$ a constant describing the curvature of space-time.  The independent variable $t$ is the proper time, i.e.\  the time measured by any comoving observer -- for practical purposes, observers at rest with respect to the Cosmic Microwave Background (CMB).

According to special relativity, the time elapsed between two events depends on the velocity of the reference frame in which it is measured.  It is shortest for an observer whose motion connects the two events.  According to GR, an observer located inside a strong gravitational field will also measure time flowing at a slower pace than an observer not experiencing such a strong field.

Similarly, we propose the following extension to GR.  We assume that an observer located inside the Universe will measure a rate of time flow depending on the state of the Universe, and thus possibly different from the time $t$ appearing in the equations of general relativity and, in particular, in the FL equation (1).  This time $\tau$, that we might call {\em cosmic time}, would be expected to flow at a variable rate as the Universe evolves.  In contrast, the {\em reference time} $t$ would be the time measured under the assumption that the rate of time flow does not vary with the evolution of the Universe, and is thus equal to the present time flow (or, by a simple change of units, to the time flow at the origin of the Universe).

Such a modification would not impact on the GR theory as long as it deals with phenomena occuring on time scales much shorter than the age of the Universe.  However, it might have a profound influence on the interpretation of phenomena occuring on cosmological times.  This theory, with a variable time flow for an observer located inside the Universe, might be called {\em universal relativity}.

\section{The flatness problem}

Space-time may have any curvature: positive for a closed, finite, Universe; negative for an open, infinite, Universe; null for a flat Universe.  In the framework of Eq.\  (1), this curvature depends on the dimensionless density parameter:
\begin{equation}
\Omega = \frac{8 \pi G \rho}{3H^2}
\end{equation} 
where $H$ is the Hubble constant:
\begin{equation}
H = \frac{1}{R} \frac{dR}{dt} \cdot
\end{equation}  
$\Omega = 1$ corresponds to a flat geometry, $\Omega < 1$ to a negative curvature and $\Omega > 1$ to a positive curvature.  Using the subscript 0 for the present value of the various quantities, it can be easily shown from the FL equation that:
\begin{equation}
\Omega = \frac{\Omega_0}{\Omega_0 + R (1 - \Omega_0)} \cdot
\end{equation}  
The present-day density parameter (considering all sources of matter and energy, including the so-called dark ones) is estimated at $\Omega_0 = 1.014 \pm 0.017$ \cite{Binney2008}.  Going back to the time when radiation decoupled from matter (giving rise to the 2.7K cosmic microwave background), i.e.\  when $R \sim 10^{-3}$, Eq.\  (4) shows that any departure from flatness would be about 1000 times smaller than at present time.  This is the so-called {\em flatness problem}: why, among all possible curvatures, is the one of the Universe so precisely equal to zero?

In the standard cosmological model, this problem is solved by inflation, which assumes that, at some very early time, a dramatic expansion occurred, annihilating any pre-existing curvature \cite{Guth1981}.  However, the physical cause of this inflation remains unknown, making it no more than an {\em ad hoc} hypothesis.

\section{Cosmic time}

An alternative hypothesis would be that an observer inside the expanding Universe experiences a cosmic time $\tau$ flowing at a variable rate, such that this observer always measures an apparent curvature of zero ($k = 0$ in the FL equation).  The FL equation would thus be written:
\begin{equation}
\left[\left(\frac{1}{R}\frac{dR}{d \tau}\right)^2 - \frac{8 \pi G \rho}{3}\right] R^2 = 0
\end{equation}
for this inside observer. 
Comparing Eqs.\  (1) and (5), it is easy to show that the unit of cosmic time $\delta \tau$ is related to the unit of reference time $\delta t$ by
\begin{equation}
\delta \tau = \frac{1}{\sqrt{\Omega}} \delta t .
\end{equation}
Equation (6) shows that, in a closed Universe $(\Omega > 1)$, the cosmic time flows faster than the reference time while, in an open Universe $(\Omega < 1)$, the cosmic time flows more slowly.  In particular, time does not flow in an empty Universe $(\delta \tau \to \infty$  when $\Omega \to 0)$.  It might be said that it is the presence of matter in the Universe that gives rise to time.

From Eqs.\  (3) and (6), it can be seen that the Hubble constant $H'$, as measured by an observer inside the Universe,  is related to the Hubble constant $H$ (measured in the reference time frame) by
\begin{equation}
H' = \frac{H}{\sqrt{\Omega}} \cdot 
\end{equation}
From (2) and (7), it is easily shown that the density parameter $\Omega$, measured in reference time, is related to the apparent density parameter ${\Omega}'$ (measured in cosmic time) by
\begin{equation}
\Omega ^2 = {\Omega}' .
\end{equation}
A popular estimate of the current (baryonic) matter density is the one deduced from Big Bang nucleosynthesis \cite{Schramm1998}.  It amounts to ${\Omega'}_0 \simeq 0.04$, thus giving $\Omega_0 \simeq 0.20$ and $\delta \tau_0 \simeq 2.2 \delta t_0$.  However, this value is somewhat model-dependent, and may not be fully appropriate as an input to a completely new model.

Another estimate, based on an inventory of different forms of baryonic matter (stars and other compact objects, cold gas, hot intracluster medium and warm-hot intergalactic medium), gives ${\Omega'}_0 \simeq 0.020-0.025$ \cite{FukugitaPeebles,CenOstriker}.  Taking ${\Omega'}_0 \simeq 0.0225$ leads to $\Omega_0 \simeq 0.15$ and $\delta \tau_0 \simeq 2.6 \delta t_0$.  In the following, we shall consider these two alternatives.

\section{Dark matter}

The higher than expected velocities of stars in the outskirts of spiral galaxies gave rise to the hypothesis that these galaxies are embedded in dark matter halos, which would have a mass of at least 4--5 times the mass of the visible matter.  Even more dramatically, the velocities of individual galaxies in galaxy clusters led to postulate that, for these clusters to remain bound, an even larger fraction of their mass must be dark \cite{Zwicky1933,Girardi2000}.  A comparison with the predictions from primordial nucleosynthesis suggests that most of this dark matter is not made of ordinary material (baryons) but of some exotic particles, with a ratio of dark to baryonic matter $\sim$6 \cite{Mushotzky2004}.  However, despite more than 20 years of investigations, these enigmatic particles have not been identified yet.

Our theory allows another interpretation of these excess velocities.  Indeed, in an open Universe, cosmic time slows down with expansion.  It means that the velocity of an object, as measured in cosmic time units, will slowly increase in the absence of any force acting on it.

Taking the velocity transformation equation from special relativity, if an object moves with a velocity $v_1$ at time $\tau_1$, an observer at time $\tau_2$ will measure a velocity $v_2$ given by:
\begin{equation}
v_2 = v_1 \frac{\delta \tau_2}{\delta \tau_1} \frac{1}{1+\frac{v_1^2}{c^2}\left( \frac{\delta \tau_2}{\delta \tau_1}-1\right)}
\end{equation} 
which reduces, for velocities much smaller than the speed of light, to:
\begin{equation}
v_2 \simeq v_1 \frac{\delta \tau_2}{\delta \tau_1} \hspace*{10pt} (v_1, v_2 \ll c) .
\end{equation}

As cosmic time varies on time scales of the order of the age of the Universe, such an effect will be completely negligible in usual experiments.  However, in galaxy clusters, where the crossing time is measured in billions of years, it may have a noticeable effect on the velocity dispersion of galaxies.

To test this, we have simulated simple galaxy clusters dynamics.  Galaxies are approximated by point masses, which move under the influence of their gravitational interactions plus that of a gas component distributed as an isothermal sphere of scale proportional to the cluster instantaneous effective radius.  The initial positions of the galaxies are chosen randomly in a sphere of a given radius.  Their initial masses are also chosen at random between a lower and an upper limit, with a uniform logarithmic distribution.  Some dynamical friction is added, and the galaxies are allowed to loose a fraction of their mass into the intracluster medium.  If the distance between two galaxies decreases below a certain (mass-dependent) threshold, galactic fusion occurs.  Galaxy clusters are assumed to form in a redshift range $6 < z < 10$.  The cluster dynamics is computed both with constant time flow and with a cosmic time variation corresponding to $\Omega_0 = 0.15 - 0.20$.  The results are then compared.

In the standard cosmology, for clusters of $\sim \!\! 400$ galaxies in a sphere of 4 Mpc radius, initial galactic masses in the range $2 \cdot 10^{9} - 4 \cdot 10^{12} \, {\rm M}_\odot$, and an initial mass of intracluster gas of the order of the total mass in galaxies, we find dynamical masses, derived from the virial theorem, ranging between 1.5 and 2.2 times the actual mass in gas and galaxies, i.e.\  a modest overestimate which may be explained by the effect of galactic fusions, the loss of galactic gas into the intracluster medium and by the fact that the clusters are not fully relaxed.  In contrast, in our cosmological model, the overestimate amounts to a factor ranging between 4 and 8, fully compatible with the total-to-baryonic mass ratio usually found in galaxy clusters \cite{Mushotzky2004}.  We can thus conclude that, in our universal relativity model, no dark matter is needed to explain the observed velocity dispersions in galaxy clusters.

\section{Evolution of the scale factor}

The evolution of the scale factor $R$ as a function of cosmic time $\tau$ may be obtained by solving Eqs.\  (1) and (6) with the appropriate value of the present matter density.  The results are displayed on Fig.\  1, which shows that our model with $\Omega_0 = 0.15-0.20$ gives results quite similar to the standard $\Lambda$CDM model, except for the first epochs $(R < 0.3$, i.e. $z > 2)$.  In particular, our model explains the present apparent acceleration of the expansion.\vspace{0.5em}
\begin{figure}[h]
\label{Images}
\includegraphics[width=0.9\textwidth]{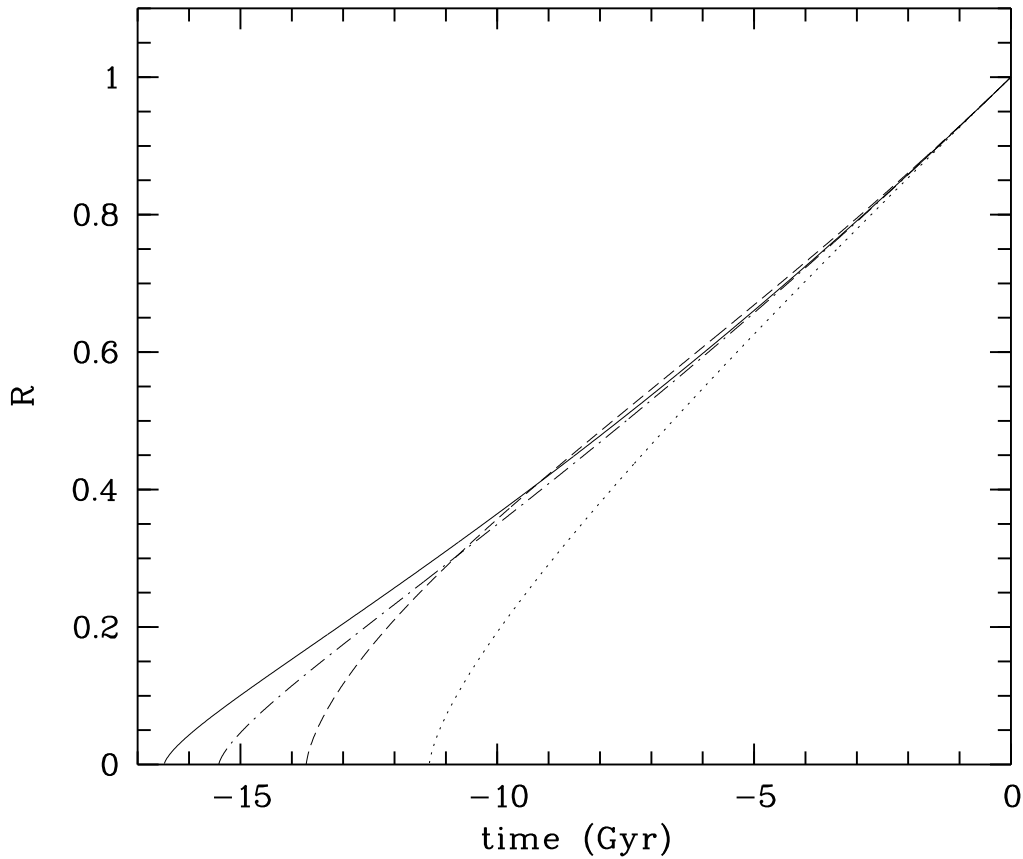}
\caption{Evolution of the scale factor predicted by 4 different cosmological models sharing the same value of the present day (apparent) Hubble constant, $H'=$ 71 km/s/Mpc: our model with $\Omega_0 = 0.15$ (full curve) and $\Omega_0 = 0.20$ (dash-dotted curve), the standard $\Lambda$CDM model (dashed curve) and a model with dark matter $(\Omega_0 = 0.27)$ and no dark energy (dotted curve).}
\end{figure}

With the universal relativity model, depending on the value chosen for $\Omega_0$, we obtain an age of the Universe of 15.4 to 16.5 Gyr, which is significantly larger than the 13.7 Gyr derived from the standard model.  In particular, the time elapsed before $z = 10$, which is about the time when the youngest galaxies are presently observed \cite{Zheng2012}, amounts to 1.0 -- 1.5 Gyr, while it is only 0.5 Gyr in the standard model.  Since the short time left for galaxy formation is a potential problem in the $\Lambda$CDM model, our cosmology allows to significantly relax this constraint.

If the evolution of the scale factor predicted by our model were to be interpreted in the framework of a model with constant time flow plus dark energy (in the form of a cosmological constant), one can show that the density parameter $\Omega$ and the dark energy parameter $\Lambda$ would satisfy the flatness criterion:
\begin{equation}
\Omega + \Lambda = 1
\end{equation}
in agreement with the results from high redshift supernov\ae\  and from the CMB.

Moreover, if such a $(\Omega,\Lambda)$ model is fitted on the scale factor curve from our cosmological model in the redshift range $0 < z < z_{\rm max}$, one obtains, for $1.0 < z_{\rm max} < 1.5$, a value of the dark energy parameter $0.70 < \Lambda_0 < 0.79$.  Thus, fitting an $(\Omega,\Lambda)$ model onto our model in the redshift range explored by the supernov\ae\  studies gives density and dark matter parameters in agreement with those deduced from such studies and from the analysis of CMB anisotropies ($\Omega_0 = 0.27$, $\Lambda_0 = 0.73$).

\section{Distant Supernovae}

The redshift-distance relation is the primary observational test for the evolution of the scale factor, and precisely the one which allowed, a decade ago, to discover the acceleration of the expansion \cite{Riess1998,Perlmutter1999}.  It is based on the use of Type Ia Supernov\ae\  (SNIa) as standard candles for distance measurements.  Two large collaborations are active in these SNIa analyses: the Supernova Cosmology Project \cite{SCP} and the High-z Supernova Search Team \cite{Davis2007,Riess2007,Wood2007}.  These teams make special effort to detect and analyze distant supernov\ae, and have been able to detect a significant amount of such objects at $z > 1$.

The Hubble diagram built from these SNIa data is compared to predictions of different cosmological models on Fig.\  2.  This figure shows that, while in the standard $\Lambda$CDM model, the inclusion of dark energy is absolutely essential to fit the data, our model -- which does not need dark matter nor dark energy -- provides an adequate fit, especially for $\Omega_0 = 0.15$, i.e.\  the lower limit of the density range considered.  We note, however, that the correction of systematic errors in the SNIa data is an extremely difficult task, and that remaining errors of the order of the difference between our model and the standard model (i.e.\  $\sim \!\! 5\%$) cannot be excluded.  In any case, observations of supernov\ae\  with $z > 2$ would allow to discriminate between the $\Lambda$CDM and the universal relativity models.
\begin{figure}[h]
\label{SN}
\includegraphics[width=0.9\textwidth]{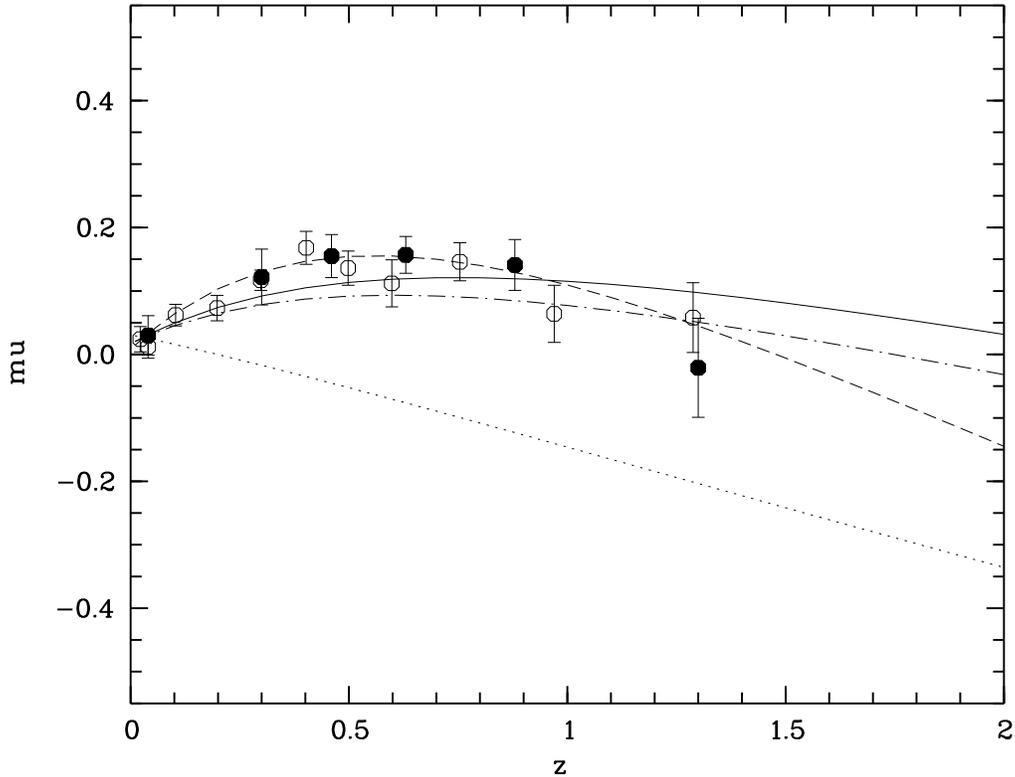}
\caption{Distance modulus $\mu$ (in magnitude units) versus redshift, after subtraction of the predictions of an empty cosmological model.  The data are from the compilations of the Supernova Cosmology Project (open circles) and of the High-z Supernova Search (filled circles), averaged over redshift bins.  The error bars are the standard errors of the mean.  The data are compared to the predictions of our model (full curve, $\Omega_0 = 0.15$; dash-dotted curve, $\Omega_0 = 0.20$), of the standard $\Lambda$CDM model (dashed curve) and of a model with dark matter $(\Omega_0 = 0.27)$ and no dark energy (dotted curve).  All models are normalized to the SNIa data in the lowest redshift bin.}
\end{figure}

\section{The horizon problem}

The horizon problem may be summarized in the following way.  Observations of the CMB show it to be extremely uniform, with temperature inhomogeneities not exceeding $\delta T / T \sim 10^{-4}$.  However, in the standard cosmological model, two points separated by more than a few degrees on the sky have never been in causal contact before the decoupling of matter and radiation.  How is it then possible to explain such a uniformity?

In the standard model, this problem is solved by inflation, which postulates an early sudden expansion of the Universe by many orders of magnitude, implying a possibility for causal contact prior to this phase.

In our model, this horizon problem may be solved quite naturally.  In the early phases of the expansion, regions of the Universe which are not in causal contact would behave independently of each other.  They would thus have their own cosmic time.  According to Eq.\  (6), if a region has a density higher than average, its cosmic time will flow faster than average.  This region will thus expand faster and, consequently, its density will drop faster.  Similarly, regions of lower density will expand more slowly.  This will ensure homogeneisation and subsequent expansion at the rate corresponding to the mean density of the Universe, without any need for causal contact.

\section{Conclusion}

From the simple assumption that the observers located inside the Universe measure a time flow such that the Universe always appears flat to them, we derive a cosmological model which allows to explain the velocity dispersion of galaxies in clusters without introducing any dark matter, as well as to explain the accelerated expansion of the Universe without any dark energy.  It also solves the flatness and horizon problems without any need for inflation.  Moreover, our model predicts a higher age of the Universe, i.e.\  between 15.4 and 16.5 Gyr, depending on the actual value of the matter-energy density.  This higher age allows to relax the tight constraints on the formation of stars and galaxies, since the age of the oldest of these objects is estimated to be over 13 Gyr, which is uncomfortably close to the 13.7 Gyr obtained from the standard $\Lambda$CDM cosmological model.

Of course, our model has to be tested against other observations, such as the flat rotation curves of spiral galaxies or the anisotropies of the CMB.  Gravitational lensing results should also be tested, and especially the masses of galaxies and galaxy clusters derived from lensing of background sources.

We stress that our model relies on a single assumption and has no free parameters.  Its inputs are the present day Hubble constant and matter-energy density, which are not adjusted but determined from observations.  It is thus much more economical than the standard model with its several adds-on (dark matter to explain the dynamics of large scale structures, dark energy to explain the accelerated expansion and inflation to solve the flatness and horizon problems).

\begin{acknowledgments}
We wish to thank Virginie Chantry, Michaël Gillon, Emmanuël Jehin and Valérie Van Grootel for fruitful discussions.  This work has been supported by ESA and the Belgian Science Policy Office under PRODEX program 90312.
\end{acknowledgments}

\end{document}